\documentclass{aastex}
\usepackage{spr-astr-addons}
\usepackage{natbib}
\usepackage{graphicx}

\RequirePackage{color}

\newcommand{\ergs}{\ifmmode {\rm erg\ s}^{-1} \else erg s$^{-1}$\ \fi}
\newcommand{\kms}{\ifmmode {\rm km\ s}^{-1} \else km s$^{-1}$\ \fi}

\begin{document}

\title{Fraction of the X-ray selected AGNs with optical emission lines in galaxy groups}
\slugcomment{To be submitted for ApSS.}
\shorttitle{Fraction of the X-ray selected AGNs with optical emission lines in galaxy groups}
\shortauthors{F. Li et al.}

\author{Feng Li\altaffilmark{1,2}, Qirong Yuan\altaffilmark{1}, Weihao Bian\altaffilmark{1}, Xi Chen\altaffilmark{1}, Pengfei Yan\altaffilmark{3}}

\email{$^\star$yuanqirong@njnu.edu.cn; $^\dag$whbian@njnu.edu.cn}

\altaffiltext{1}{Department of Physics and Institute of Theoretical Physics, Nanjing Normal University, Nanjing 210023, China}
\altaffiltext{2}{School of Mathematics and Physics, Changzhou University, Changzhou 213164, China}
\altaffiltext{3}{School of Mathematics and Physics, Qingdao University of Science and Technology, Qingdao 266061, China}

\begin{abstract}

Compared with numerous X-ray dominant active galactic nuclei (AGNs) without emission-line signatures in their optical spectra, the X-ray selected AGNs with optical emission lines are probably still in the high-accretion phase of black hole growth. This paper presents an investigation on the fraction of these X-ray detected AGNs with optical emission-line spectra in 198 galaxy groups at $z<1$ in a rest frame 0.1-2.4 keV luminosity range $41.3 < {\rm log}(L_{\rm X}/{\rm erg}~{\rm s}^{-1}) < 44.1$ within the Cosmological Evolution Survey (COSMOS) field, as well as its variations with redshift and group richness. For various selection criteria of member galaxies,  the numbers of galaxies and the AGNs with optical emission lines in each galaxy group are obtained. It is found that, in total 198 X-ray groups, there are 27 AGNs detected in 26 groups. AGN fraction  is on everage less than $4.6 (\pm 1.2)\%$ for  individual groups hosting at least one AGN. The corrected overall AGN fraction for whole group sample is less than $0.98 (\pm 0.11) \%$. The normalized locations of group AGNs show that 15 AGNs are found to be located in group centers, including all 6 low-luminosity group AGNs ($L_{\rm 0.5-2 keV} < 10^{42.5} {\rm erg}~{\rm s}^{-1}$). A week rising tendency with $z$ are found: overall AGN fraction is 0.30-0.43\% for the groups at $z<0.5$, and  0.55-0.64\% at $0.5 < z < 1.0$. For the X-ray groups at $z>0.5$, most member AGNs are X-ray bright, optically dull, which results in a lower AGN fractions at higher redshifts. The AGN fraction in isolated fields also exhibits a rising trend with redshift, and the slope is consistent with that in groups. The environment of galaxy groups seems to make no difference in detection probability of the AGNs with emission lines. Additionally,  a larger AGN fractions are found in poorer groups, which implies that  the AGNs in poorer groups might still be in the high-accretion phase, whereas the AGN population in rich clusters is mostly in the low-accretion, X-ray dominant phase.

\end{abstract}

\keywords{galaxies: active -- galaxies: nuclei -- galaxies: groups -- galaxies: field -- quasars: emission lines}

\section{Introduction}

It is believed that Active galactic nuclei (AGNs) are powered by the accretion around the central  supermassive black holes (SMBHs). The trigger mechanism of such nuclear activity is still subject to debate. The co-evolution between the growth of SMBHs and the formation of stars in galaxies suggests that processes to trigger star formation also trigger SMBHs accretion in AGNs \citep[e.g.,][]{Boyl98,Fran99,Fer-Mer00,Gebh00,Trem02,Merl04,Silv08, Mart09,Mart13,Kor-Ho13}.
It is possible that SMBHs accretion in luminous AGNs are triggered by major mergers of gas-rich galaxies \citep[e.g.,][]{Kau-Hae00,Sand88,Bar-Her91,Hopk06,Hop-Qua11}, which is supported by the small-scale excess of quasar-quasar and quasar-galaxy pairs \citep[][]{Henn06,Serb06}. For less luminous AGNs, some other mechanisms, such as minor mergers, large-scale bars, disk instabilities, turbulence in the interstellar medium (ISM), have also been proposed to fuel star formation and SMBH growth at lower rates \citep[e.g.,][]{Simk80,Elme98,Genz08,Hop-Qua10,Hop-Qua11}.

Groups of galaxies are supposed to be an ideal environment for such gas-fueling mergers due to their high galaxy densities and low velocity dispersions. However, for the denser environment, i.e., rich clusters of galaxies, their hotter intracluster media (ICM) and larger velocity dispersions may drive some additional processes to impact the cold gas fueling of the SMBHs \citep[e.g.,][]{Miya11,Bekk14,Bose14,Koul14, Vij-Ric15,Peng15}. These processes include the removal of cold gas via ram pressure stripping, evaporation of galactic gas by the hot ICM, tidal effects due to the cluster potential, and gas starvation due to the absence of new infall of cold gas.

The AGN fraction in galaxy groups/clusters provides valuable additional observational constraints on AGN fueling mechanisms and the growth of the central SMBHs, as well as the impact of AGNs on the intracluster medium (ICM) over cosmic time \citep[e.g.,][]{Mart09}. Nearly complete absence of quasars in rich clusters can be interpreted by the relative lack of cold gas and major mergers \citep[][]{Bar-Her92}. Investigation on fraction of the AGNs with optical emission lines selected from the Sloan Digital Sky Survey (SDSS) show that luminous AGNs are rarer in denser environments \citep[][]{Kauf04}, while the fraction of low-luminosity AGNs in SDSS and radio observations does not decrease remarkably in richer clusters \citep[][]{Mill03,Best05,Mart06}.

Based on different AGN selection methods, such as optical spectra, X-ray luminosity, mid-infrared color index and radio luminosity, the AGN fraction within groups/clusters has been extensively studied, as well as observational properties of AGNs in groups/clusters \citep[e.g.,][]{Dres85,Mart06,Mart07,East07,Mart09,Hagg10,Tomc11,Mart13,Oh14,Tzan14}.
For a very nearby sample of 1095 galaxies in rich clusters and 173 field galaxies with optical spectra, \cite{Dres85} found that 1\% of cluster galaxies host AGNs, while 5\% of field galaxies do. With a spectroscopic survey of X-ray point sources in eight low-redshift clusters of galaxies with $0.05 < z < 0.31$, \cite{Mart06} estimated that the fraction of cluster galaxies with $M_{\rm R} < -20$ mag which host the AGNs with $L_{\rm X} > 10^{41} \ergs $ is $\sim 5\%$. For 32 galaxy clusters with $0.05 < z < 1.3$, \cite{Mart09} found the AGN fraction is 0.134\% in low-$z$ clusters ($z < 0.4$) , and increase significantly to 1\% in high-$z$ clusters ($z > 0.4$) \citep[also see][]{East07}. For a sample of 13 clusters of galaxies at higher redshifts (i.e., $1<z<1.5$), \cite{Mart13} found that the AGN fraction in clusters is $3.0_{-1.4}^{+2.4}\%$, suggesting the rising tendency of AGN fraction in clusters with redshift from 0.05 to 1.5. With the {\it Chandra} Multiwavelength Project (ChaMP) and SDSS data, using $L_{0.5-8 \rm{keV}}>10^{42} \ergs $ to select AGNs, \cite{Hagg10} found that the AGN fraction in groups is $0.16 \%$ for $z\leq 0.125$, and $3.8  \%$ for $z\leq 0.7$, depending on the selection criteria of field galaxies \citep[also see][]{Mart09}. For a mass-selected sample of galaxies from the 10k catalog of the zCOSMOS spectroscopic redshift survey, \cite{Silv09} identified 147 AGNs, and found no significant difference of AGN fraction in groups with that in fields.

Compared with optically selected AGNs, the AGNs revealed by $Chandra$ and $XMM-Newton$ surveys have a higher spatial number  density that peaks at low redshift \citep[][]{Hasi05}, and show a stronger spatial clustering \citep[][]{Yang06}.The fraction of X-ray selected AGNs in clusters is significantly higher than that of the optically selected AGNs with emission lines \citep[e.g.][]{Mart09}. Large number of X-ray AGNs in clusters show little or no optical AGN signatures \citep[][]{Mart06}. \cite{Shen07} studied the AGN population in eight low-redshift poor groups of galaxies, and found that the X-ray bright, optically dull AGNs are entirely absent from the less  dynamically evolved groups, which supports a  scenario for AGN accretion evolution: AGN activity is initially triggered by galaxy merging, leading to a high accretion rate and an optically dominant phase. With the decline of accretion rate, the AGN gradually enters into an X-ray dominant phase without optical emission lines.  Thus the optically and X-ray selected AGNs harbor the same population of SMBHs observed at different epochs. According to this picture, the majority of AGNs in poor galaxy groups are likely in the high-accretion phase with optical emission signatures, and the early phase of accretion evolution should be more sensitive to the ideal environment of gas-fueling mergers --- groups of galaxies.

In this paper, we plan to take a large sample of X-ray selected galaxy groups in the COSMOS field to investigate the fraction of X-ray selected AGNs with optical emission lines which might be in the high-accretion phase.  This investigation will focus on cosmological evolution and richness dependence of the AGN fraction in the redshift region $z<1$. \S 2 describes the data and samples . The AGN fractions in groups and in fields are given in \S 3, as well as  its cosmological evolution and variation with group richness. The main conclusions are given in \S 4. Throughout this work we assume a standard $\Lambda${\sc CDM} cosmology with $\Omega_{\rm \Lambda}=0.7$, $\Omega_{\rm M}=0.3$ and $H_{0}=70$\,\kms\,Mpc$^{-1}$.

\section{Data and sample}

\subsection{Galaxy groups in COSMOS}

The COSMOS field covers about 2 deg$^2$ equatorial area in the constellation {\it Sextans}. It has been photometrically observed in a broad range of wavelengths (over 30 bands from the ultraviolet, optical, and infrared), and it also has been mapped through 54 overlapping $XMM-Newton$ pointing and additional $Chandra$ observations for the central 0.9 deg$^2$ region \citep[e.g.][]{Hasi07, Fino07, Ilbe09, Brus10}. Group catalogs have been constructed in this field on the basis of X-ray data \citep[e.g.,][]{Fino07,Geor11},  zCOSMOS spectroscopy \citep[e.g.,][]{Knobel09}, and photometric redshifts \citep[][]{Gilli11}.

In this paper we will take the group catalog given by \cite{Geor11}, including 211 extended X-ray selected galaxy groups ($0 < z < 1$)  in a rest-frame 0.1-2.4 keV luminosity range of $41.3 < {\rm log}(L_{\rm X}/{\rm erg}~{\rm s}^{-1}) < 44.1$ in the COSMOS field, among which 165 groups have reliable optical counterparts \citep[]{Fino07,Fino10} \footnote{http://irsa.ipac.caltech.edu/data/COSMOS/tables/groups/}. Using a restrict Bayesian algorithm, \cite{Geor11}  assigned 115,844 galaxies to groups based on precise photometric redshifts \citep{Ilbe09,Alle12}. These galaxies satisfy  $z_{ph}<1.2$ and I(F814W) $< 24.2$ mag, and require a $3\sigma$ detection in the $K_s$ band for the stellar mass estimation, which is complete to a typical depth of $K_s = 24$.  With $L_{\rm X}-M$ scaling relation calibrated by weak gravitational lensing, the halo masses $M_{200}$ are calculated in the range $10^{13} \leq M_{200}/M_{\odot} \leq 10^{14}$ for these galaxy groups within $R_{200}$, where $M_{200}\equiv M(<R_{200})=200\rho_{c}(z) \frac {4}{3} \pi R_{200}^3$, and $R_{200}$ is the radius at which the mean interior density is equal to 200 times the critical density \citep{Leau10}. Roughly 20\% of group members have spectroscopic redshifts.

The membership probability for each galaxy in groups, $P_{\rm mem}$, is given by \cite{Geor11}. With a spectroscopic subsample, \cite{Geor11} estimated the purity ($p$) and completeness ($c$) for different membership probability thresholds. For a threshold of $P_{\rm mem} > 0.1$, the purity and completeness of member selection are about 63\% and 97\%, respectively.  With a higher threshold of $P_{\rm mem}$, the purity is improved, and the completeness is deteriorated.  For $P_{\rm mem} > 0.3$, $p= 67$\% and $c= 94$\%. For $P_{\rm mem} > 0.5$,  $p=69$\% and $c= 92$\%. Taking the threshold of $P_{\rm mem} > 0.1$, there are 198 X-ray selected groups with $z < 1.0$   in the COSMOS field.

\subsection{X-ray selected AGNs with emission lines}

Entire COSMOS field (2~deg$^2$) has been mapped by the $XMM-Newton$ pointing observations ($\sim 1.55$~Ms),  and only the central region (0.9~deg$^2$) is covered by the $Chandra$ observations  with a higher spatial resolution \citep[]{Fino07,Brus10}.
With the $XMM-Newton$ survey of the COSMOS field, there are 1,848 point-like sources detected, at least one of the soft (0.5-2 keV), hard (2-10 keV) and ultra-hard (5-10 keV) bands down to limiting fluxes of $5 \times 10^{-16}$, $3 \times 10^{-15}$, and $7 \times 10^{-15}$ erg~cm$^{-2}$~s$^{-1}$, respectively. Only about half sources are detected with the $Chandra$.

To match the sample of 198 X-ray selected groups  with $z<1$, we need to construct a contiguous (2 deg$^2$) X-ray selected AGN sample, with well-defined flux limits and a reliable estimate of spectroscopic completeness.  In this paper we will take the AGN sample obtained from the medium-depth ($\sim$ 60 ks) $XMM-COSMOS$  survey. \cite{Brus10} matched a multi-wavelength counterpart to 1,797/1,848 sources\footnote{http://vizier.cfa.harvard.edu/viz-bin/VizieR?-source=J/ApJ/716/348}.  There are 730 AGNs identified by a lot of good-quality spectroscopic follow-up programs of XMM point sources, such as  Magellan/IMACS and MMT observation campaigns, and VIMOS/zCOSMOS bright and faint projects \citep[e.g.][]{Gilli09}.  The objects having at least one broad optical emission line (FWHM $> 2000$ \kms) are classified as broad-line AGNs. Those with unresolved high-ionization lines, or exhibiting line ratios indicating AGN activity in the BPT diagram \citep[][]{Bald81} are classified as narrow-line AGNs. For the remaining objects whose spectral range does not allow to construct line diagnostics, the objects with rest-frame hard X-ray luminosities in excess than $2\times 10^{42}$ \ergs are also classified as narrow-line AGNs \citep[see section 5.1 in][]{Brus10}. The majority (403, $\sim 55\%$) are broad-line AGNs, whereas the rest AGNs are narrow-line ones. Majority of broad-line AGNs are observed at $z>1.0$ and with  0.2-2 keV X-ray luminosities  ${\rm log}(L_{\rm X} > 41.5\, {\rm erg}~{\rm s}^{-1}) $, while narrow-line AGNs are observed down to very low X-ray luminosities \citep[see Figure 2 in][]{Gilli09}. \cite{Gilli09}  studied the spatial clustering the X-ray selected AGNs in the XMM-Newton field, based on an analysis of the sample of 538 X-ray selected AGNs with $I_{\rm AB} < 23$ and  $0.2 < z < 3.0$. They estimated the spectroscopic completeness is about 60\% for $z<1.2$ and the spectroscopic selection does not include any bias against high-redshift objects. Following \cite{Gilli09}, we take $I_{\rm AB} < 23$ and $ z < 1.0$,  338 out of 730 AGNs ($\sim 46.3\%$) are selected to have optical emission line signatures. This spectroscopic completeness  ($\sim$60\%) will be adopted in our further estimate of AGN fraction.

\subsection{AGNs/galaxies in groups and fields }

\begin{figure}[htbp]
\center
\includegraphics[width=6cm,height=9.15cm]{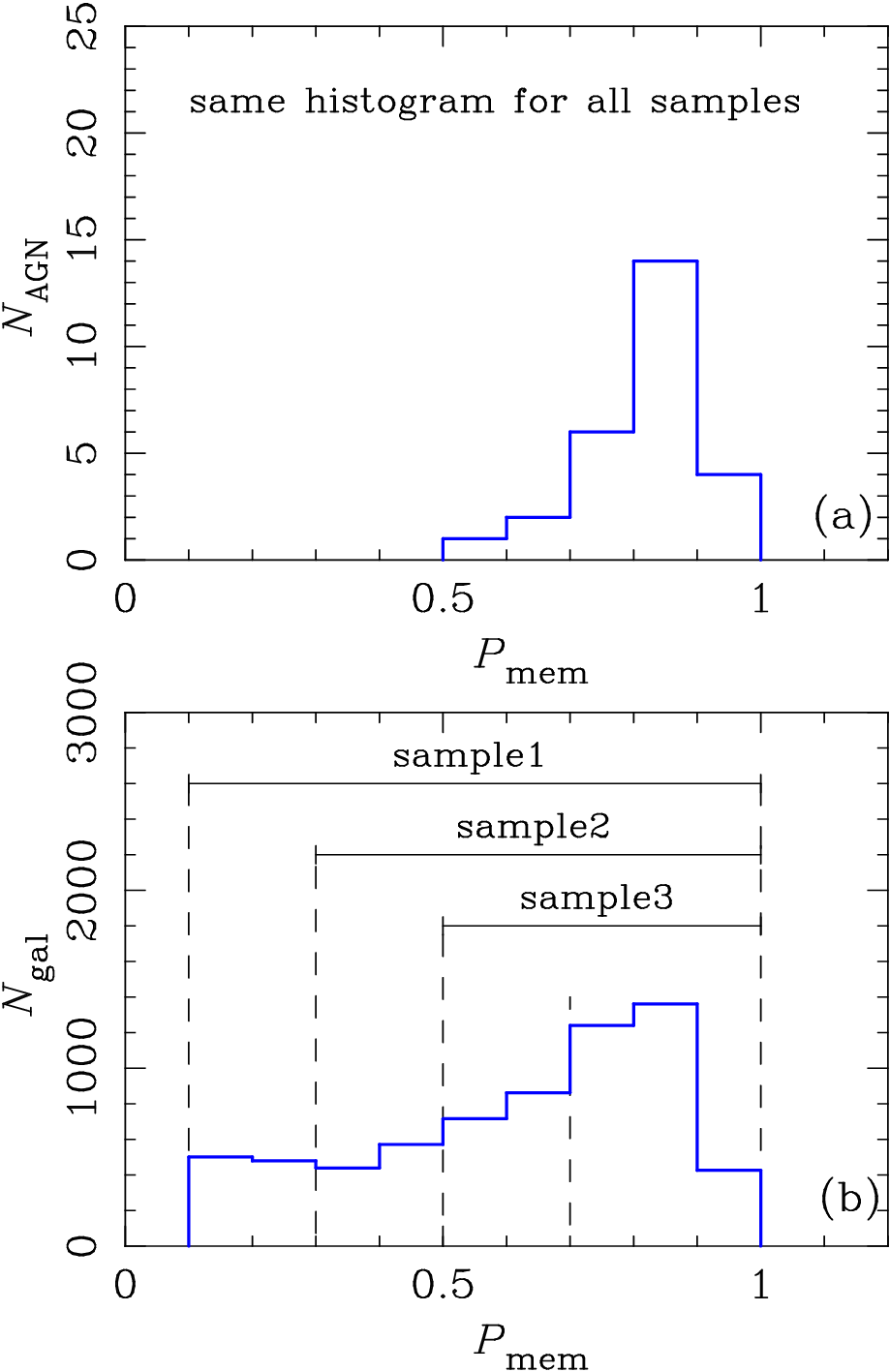}
\caption{Distribution of membership probability ($P_{\rm mem}$) for 27 AGNs and 6,597 galaxies in groups with $P_{\rm mem}>0.1$.}
\label{ffig1}
\end{figure}

Given a threshold of the membership probability $P_{\rm mem}$, the member galaxies for each group are obtained. Then we cross-identify above AGN sample with the member galaxies. With $P_{\rm mem} > 0.1$, there are 27 AGNs in 26 groups with 6,597 member galaxies (see Table~\ref{table1}), and their distributions of membership probabilities, $P_{\rm men}$ are shown in Fig.~\ref{ffig1}. For these 27 AGNs in 26 galaxy groups, the AGN membership probability ($P_{\rm mem}$) are large than 0.5, with a peak in the range of $0.8 < P_{\rm mem} < 0.9$. The 27 AGNs in groups are nearly all narrow-line AGNs, except one broad-line AGN, which can be interpreted by the fact that the broad-line AGNs are on average high-luminosity AGNs at $z>1$.   For selecting field galaxies with $z<1$, we take $P_{\rm mem} = 0$ and GROUP\_FLAG $=-1$, and obtain 74,350 field galaxies, including 299 AGNs with emission lines.

\begin{figure}[htbp]
\center
\includegraphics[width=6cm,height=9.15cm]{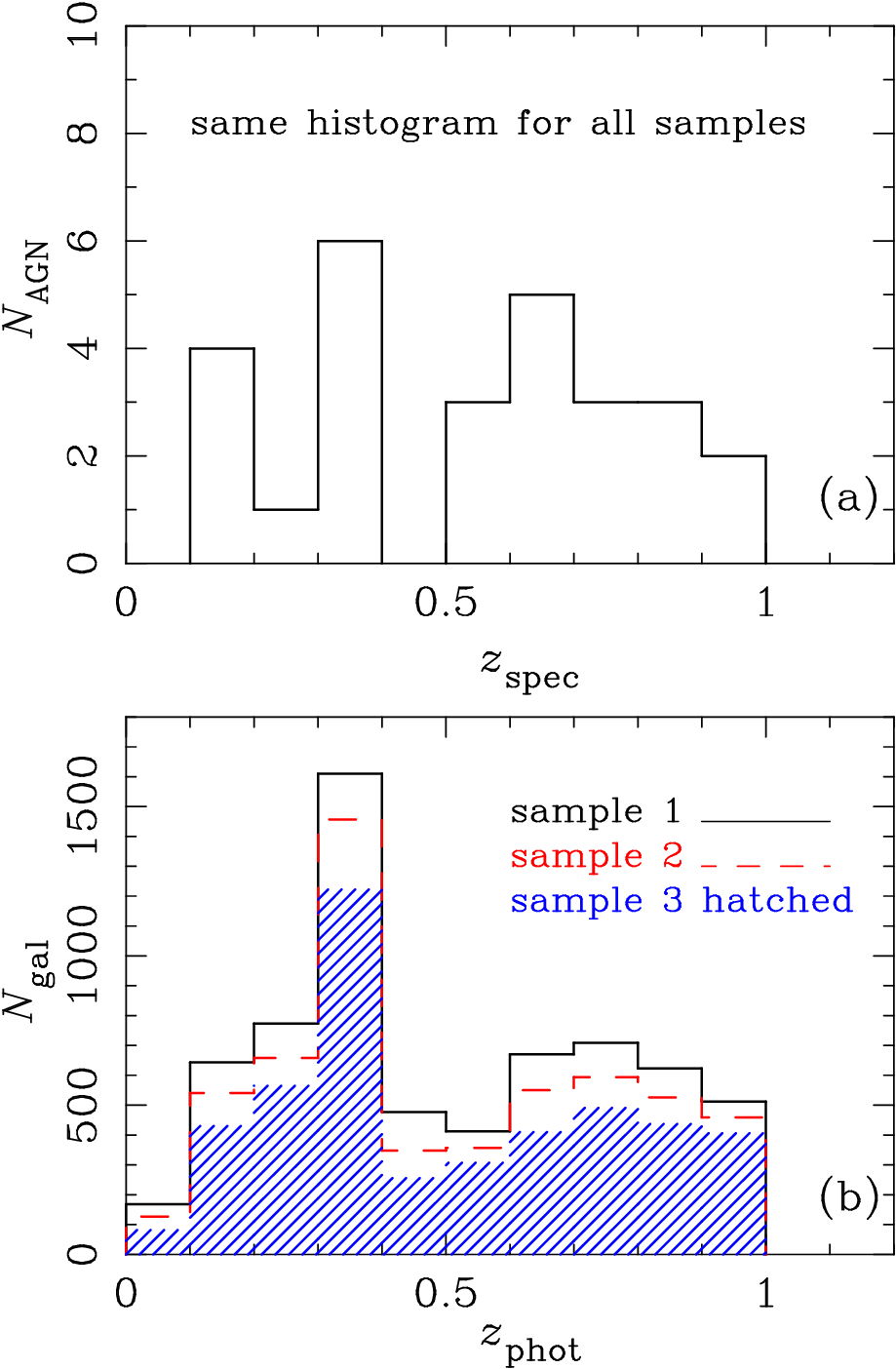}
\caption{Redshift distributions of 27 AGNs and 6,597 galaxies in groups for three $P_{\rm mem}$ thresholds. }
\label{ffig2}
\end{figure}

In order to avoid the selection effect on membership assignment, we adopt three membership probability thresholds (i.e. 0.1, 0.3 and 0.5), and  obtain three samples of member galaxies/AGNs in groups, respectively. Table \ref{table1} shows the statistics of 27 AGNs in 26 groups and galaxies in all 198 groups for three $P_{\rm mem}$ thresholds, including purity, completeness, number of member galaxies and AGNs, and the overall mean AGN fraction in all 198 groups. The distributions of membership probability ($P_{\rm mem}$) for 27 AGNs and galaxies in all 198 groups are shown in Fig.~\ref{ffig1}. The redshift distributions for three samples of group members are shown in Fig.~\ref{ffig2}. It is found that the peak redshift of group galaxies appears in the range of $0.3 < z < 0.4$ (see Fig.~\ref{ffig2}). As $P_{\rm mem}$ threshold increases, the number of member galaxies decreases, but the distribution with photometric redshifts ($z_{\rm phot}$) remains unchanged. The Kolmogorov-Smirnov tests show that three $z_{\rm phot}$ distributions obey the same distribution with the probabilities more than 86\%.

\begin{center}
\begin{table*}[t]
\caption{Statistics for three samples of member galaxies/AGNs in groups.}
\label{table1}
\begin{tabular}{ccccccccc}\hline \hline

No. of  & $P_{\rm mem}$ & Number of & Number of & Number of & Purity & Completeness & $\langle f_{\rm AGN} \rangle$ & $\langle f^{\rm corr}_{\rm AGN}\rangle$ \\
Samples & Threshold  & Groups  & AGNs  & Galaxies  & ($p$)  & ($c$)  & (\%) & (\%)\\ \hline

1 & 0.1 & 26 & 27 &  6597 & 0.63 & 0.97 & $0.41 \pm 0.08$ & $0.68 \pm 0.08$\\
2 & 0.3 & 26 & 27 &  5617 & 0.67 & 0.94 & $0.48 \pm 0.09$ & $0.80 \pm 0.09$\\
3 & 0.5 & 26 & 27 &  4606 & 0.69 & 0.92 & $0.59 \pm 0.11$ & $0.98 \pm 0.11$\\ \hline

\end{tabular}
\end{table*}
\end{center}

\section{AGN fraction in groups and fields}

\subsection{The AGN fraction in each group }

\begin{figure*}[htbp]
\center
\includegraphics[width=17cm,height=4.38cm]{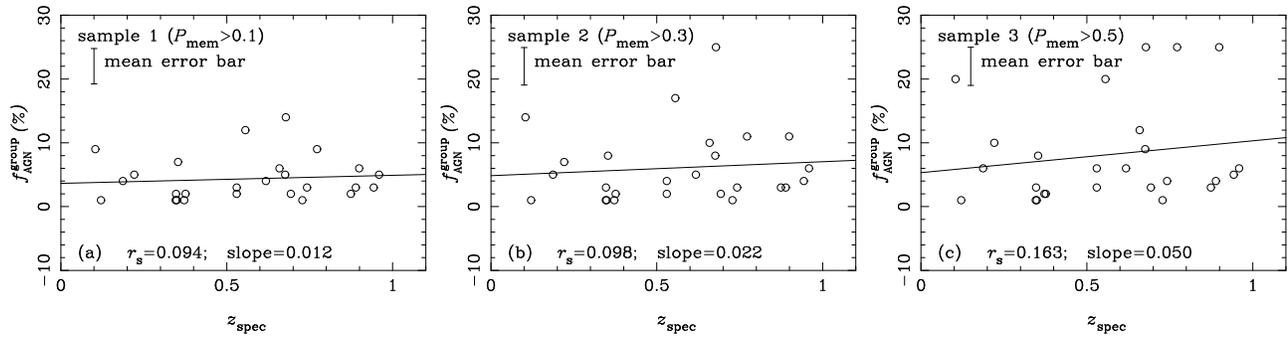}
\caption{The AGN fraction versus as the AGN spectroscopic redshift for three samples of group galaxies. The solid line is the linear fit. The Spearman correlation coefficient ($r_s$) and the slope of the best linear fit are presented in the bottom in each panel, as well as the mean error on the left corner.}
\label{ffig4}
\end{figure*}

From 198 X-ray selected groups ($z < 1.0$, $P_{\rm mem} > 0.1$) in the COSMOS field, it is found that there are 27 AGNs appearing in 26 groups. The AGN fraction for a given group is defined as $f^{\rm group}_{\rm AGN} = N^{\rm group}_{\rm AGN}/N^{\rm group}_{\rm gal}$, where $N^{\rm group}_{\rm AGN}$ and $N^{\rm group}_{\rm gal}$ are the numbers of AGNs and galaxies in a group, respectively. We calculate the AGN fractions in these 26 groups, as well as their errors.  We estimate 1 $\sigma$ errors based on a Poisson distribution due to the small number of objects in each group \citep[]{Silv08}. Considering the purity ($p$) and completeness ($c$), the AGN fraction ($f^{\rm group}_{\rm AGN}$) keeps unchanged, the corresponding error will be magnified by $\sqrt {c/p}$. Fig.~\ref{ffig4} shows the the AGN fraction in each group as a function of spectroscopic redshifts  for these 26 groups. The Pearson correlation coefficient ($r_{s}$), and the slope ($\Delta f^{\rm group}_{\rm AGN}/\Delta z$) of the linear fitting are also presented. It can be seen that, for three samples, $f^{\rm group}_{\rm AGN}$ has a very weak rising trend with redshift, and the linear correlations are very weak. For quantifying the distribution of AGN fractions in groups, we apply
the ROSTAT software \citep[][]{BFG90} to calculate the biweight location, which is analogous to the mean value. The biweight locations of AGN fraction for three samples are $3.5(\pm0.6)\%$, $4.5(\pm0.9)\%$, $4.6(\pm1.2)\%$, respectively, suggesting that the average AGN fraction in X-ray groups at $z < 1$ is less than 5\% for individual groups.

It should be noted that, for more than 87\% of X-ray groups, none of member galaxies has been spectroscopically identified as AGNs.  Table \ref{table1} shows the overall mean AGN fraction for three samples. For larger probability threshold, number of group galaxies is smaller, and the AGN number remain the same, resulting in larger overall mean AGN fraction, $\langle f_{\rm AGN} \rangle$, from 0.41\% to 0.59\%. Considering  the spectroscopic completeness of about 60\% \citep[][]{Gilli09},  the mean AGN fraction can be corrected for each sample. The corrected mean AGN fractions, $\langle f^{\rm corr}_{\rm AGN}\rangle$ , are listed in Table \ref{table1}. Conservatively, for 198 X-ray selected groups at $z<1$, overall fraction of the AGNs with emission lines is less than 1 \%.

\subsection{Location of 27 AGNs in 26 galaxy groups}

\begin{figure}[htbp]
\center
\includegraphics[width=6cm,height=9.25cm]{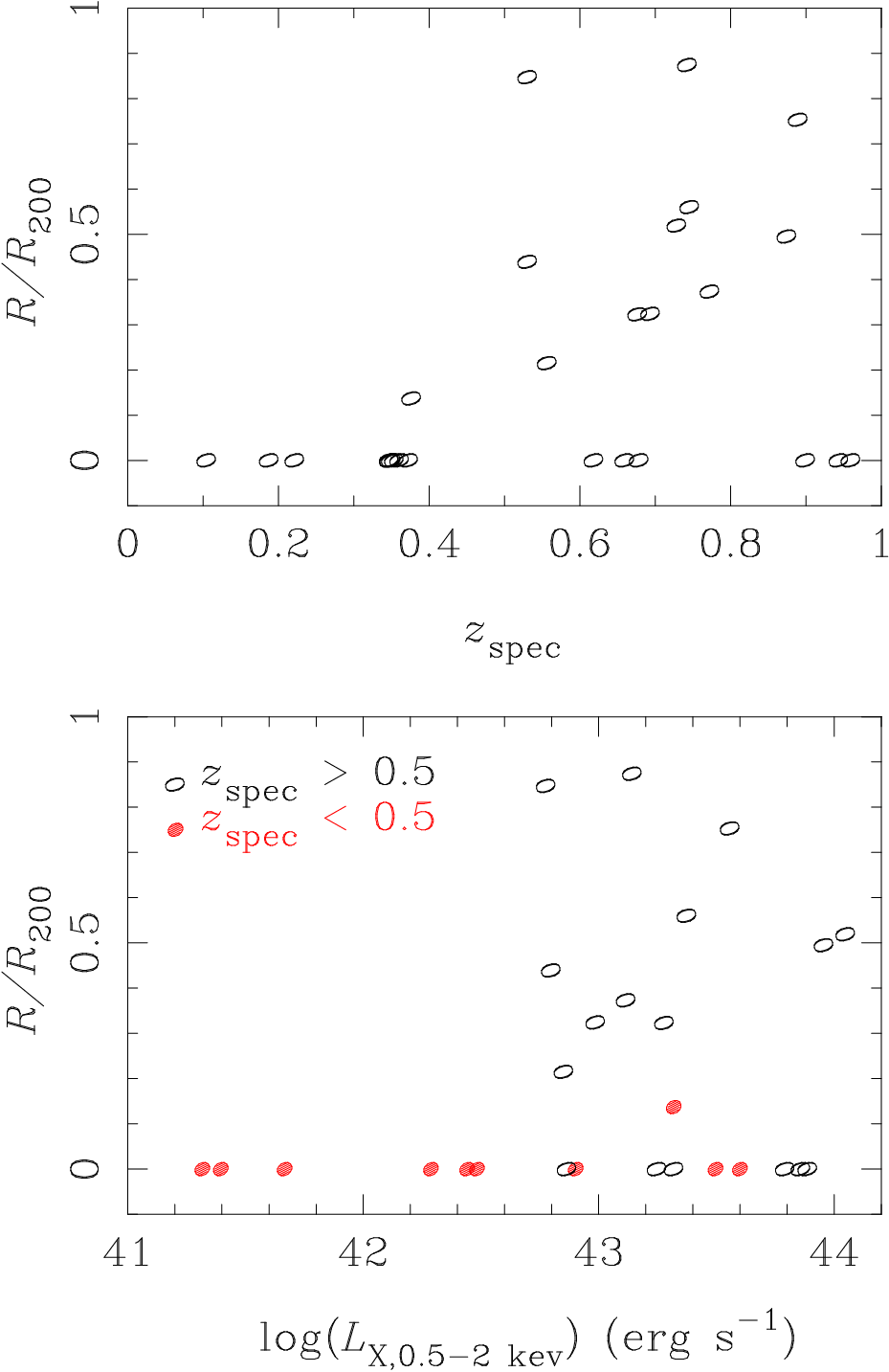}
\caption{Distribution of normalized projected distance, $R/R_{200}$, for 27 AGNs as a function of spectroscopic redshift in groups (top) and the X-ray rest-frame luminosity in soft band (bottom).}
\label{ffig3}
\end{figure}

It is hard to investigate the distribution of X-ray luminous AGNs within individual groups/clusters due to the rarity of luminous AGNs and large uncertainties of photometric redshifts. Fig.~\ref{ffig3} shows normalized projected distance to the most massive galaxy in each group, $R/R_{200}$,  of 27 AGNs versus spectroscopic redshifts and the rest-frame soft X-ray luminosity ($0.5 - 2$~keV). There are 15 out of 27 AGNs with $R/R_{200} = 0$, suggesting that they are all the most massive group galaxies. For the groups  at $z < 0.5$, 9 out of 10 AGNs are located at group center (i.e., $R/R_{200} = 0$). For  the groups at $z>0.5$, some emission-line AGNs are located in outer region of groups, up to $\sim 0.9 R_{200}$. There are 21 out of 27 AGNs with $L_{\rm X, 0.5-2 keV} > 10^{42.5}$ erg s$^{-1}$, indicating that majority of the group AGNs are X-ray luminous galaxies. It is interesting to see that all 6 faint group AGNs with $L_{\rm X, 0.5-2 keV} < 10^{42.5}$ erg s$^{-1}$ are found to be located at group centers. The low-luminosity AGNs seems to have higher probabilities to be found at the centers of low-$z$ groups of galaxies.

\subsection{Cosmological evolution of overall AGN fraction}

\begin{figure*}[thbp]
\center
\includegraphics[width=17cm,height=4.2cm]{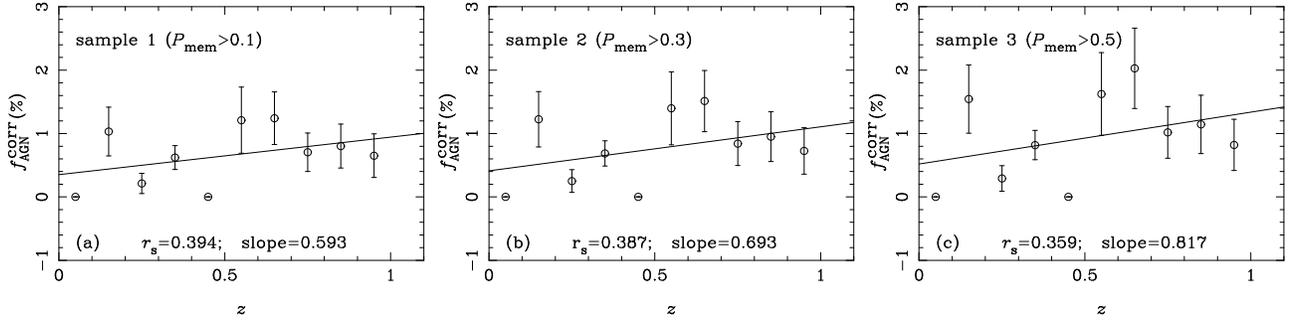}
\caption{Overall AGN fraction versus the median redshift in 10 $z$-bins for 198 X-ay selected galaxy groups. The solid line is the linear fit. The Spearman correlation coefficient ($r_s$) and the slope of the best linear fit are presented in the bottom in each panel.}
\label{ffig5}
\end{figure*}

For $P_{\rm mem} > 0.1$, there are 26 X-ray selected  groups at $z<1$ hosting at least one AGN with emission lines. For the rest of groups, no  galaxies are found to be in the high-accretion AGN phase. The groups with no detection of emission-line AGNs  should be taken into account when we evaluate the overall AGN fraction in groups.  In order to investigate cosmological evolution of overall AGN fraction in groups, the sample of 198 X-ray groups are divided into 10 $z$-bins with a interval width of $\Delta z = 0.1$. Considering the spectroscopy completeness, we calculate the overall AGN fraction in groups for each $z$-bin, $f^{\rm corr}_{\rm AGN} = N_{\rm AGN}/(0.6N_{\rm gal})$, where $N_{\rm AGN}$ and $N_{\rm gal}$ are the number of emission-line AGNs and galaxies in groups for each $z$-bin, respectively. Due to small number of objects in each $z$-bin, 1 $\sigma$ errors of  number counts are estimated  on the basis of a Poisson distribution \citep[][]{Silv08}. The error of overall AGN fraction in each $z$-bin can be  derived by error-propagation, and  considering a magnifying factor of $\sqrt{c/p}$. As illustrated in Fig. \ref{ffig5}, overall AGN fraction ($f^{\rm corr}_{\rm AGN}$)  is found to have a slight rising trend with redshift, and their Spearman correlation coefficients, $r_{s}$,  are in a range from 0.36 to 0.39 for three samples. The rising tendency for overall AGN fraction is basically consistent with the results in \cite{Mart09} and \cite{Oh14}.

Because of small number of group AGNs with emission lines, the tendency of overall AGN fraction is rather uncertain for the binning strategy of $\Delta z = 0.1$,  For decreasing uncertainties in AGN fraction estimates, we split samples into low-$z$  ($z < 0.5$) and high-$z$  ($0.5 < z < 1.0$) subsamples. For sample 1 ($P_{\rm mem} > 0.1$),  overall AGN fractions are $0.30(\pm 0.11) \%$ and  $0.55(\pm 0.17) \%$ for low- and high-$z$ bins, respectively. For sample 2 ($P_{\rm mem} > 0.3$), they are $ 0.35(\pm 0.13) \%$ and $ 0.64 (\pm 0.19) \% $ respectively for low- and high-$z$ subsamples. For sample 3 ($P_{\rm mem} > 0.5$), overall AGN fractions are $0.43(\pm 0.15) \%$ and $0.64 (\pm 0.19) \%$ . In sum, the overall AGN fraction in groups is 0.30-0.43\% at $z<0.5$, and it is 0.55-0.64\% at $0.5 <z<1.0$.

Considerable efforts have been made to measure the AGN fraction with redshifts in groups and clusters. \citet{East07} was the first to report a significant increase of AGN fraction in galaxy clusters with two luminosity thresholds ($L_{\rm X, H} > 10^{42}~\ergs$ and $L_{\rm X, H} > 10^{43}~\ergs$) from $z \sim 0.2$ to $z \sim 0.6$. \cite{Mart09} also obtained a significant increase of AGN fraction for 32 galaxy clusters with $0.05 < z < 1.3$, finding the AGN fractions are $0.134\%$ in low-$z$ clusters ($z < 0.4$) and $1.00\%$ in high-$z$ clusters ($z > 0.4$), respectively. Taking $L_{\rm X,0.5-8 {\rm keV}}>10^{42} \ergs $ to select AGNs, \cite{Hagg10} found that the AGN fraction in groups is $0.16(\pm 0.06)\%$ for $z\leq 0.125$, and $3.8(\pm 0.92)\%$ for $z\leq 0.7$. Using three different selection techniques (mid-IR color, radio luminosity, and X-ray luminosity), \citet{Gala09} extended this rising trend with redshift up to $z \sim 1.5$, and the observed increase with redshift is more pronounced in the galaxy cluster than in the field \cite[see][Figure 2]{Gala09}. Our result also presents an increasing trend of the fraction of emission-line AGNs with redshift, which is in accordance with most previous studies. This suggests the evolution of AGN population in rich clusters has a Butcher-Oemler effect similar to star forming galaxies \citep{But-Oem84}.
It should be noted that the significant variations in AGN fraction estimate among previous works are mainly due to different selection criteria of the samples of AGNs and cluster galaxies, which result in somewhat different redshift distributions and limiting magnitudes for the samples under comparison.

\begin{table*}[htp]
\centering \caption{Statistics for galaxies and AGNs in all 198 groups ($P_{\rm mem} > 0$) and in fields ($P_{\rm mem} = 0$)}
\label{table2}
\begin{tabular}{cccccccc}
\hline
\hline
redshift & Number of & Number of & Number of & Number of & Number of & $f^{\rm corr}_{\rm AGN}$ (\%) & $f^{\rm corr}_{\rm AGN}$ (\%) \\
Range    & all AGNs & group AGNs & field AGNs & group galaxies & field galaxies  & in group & in field\\
\hline
$(0.0, 0.1)$ &  3 & 0 &  3 &  232 &  1514 & $0$             & $0.33 \pm 0.11$ \\
$[0.1, 0.2)$ & 12 & 5 &  7 &  845 &  3149 & $0.99 \pm 0.27$ & $0.37 \pm 0.08$ \\
$[0.2, 0.3)$ & 12 & 1 & 11 & 1228 &  4975 & $0.14 \pm 0.08$ & $0.37 \pm 0.07$ \\
$[0.3, 0.4)$ & 41 & 7 & 34 & 1903 &  9235 & $0.61 \pm 0.14$ & $0.61 \pm 0.06$ \\
$[0.4, 0.5)$ & 21 & 1 & 20 &  782 &  7574 & $0.21 \pm 0.13$ & $0.44 \pm 0.06$ \\
$[0.5, 0.6)$ & 27 & 3 & 24 &  521 &  7239 & $0.96 \pm 0.33$ & $0.55 \pm 0.07$ \\
$[0.6, 0.7)$ & 48 & 5 & 43 &  836 & 10311 & $1.00 \pm 0.27$ & $0.70 \pm 0.06$ \\
$[0.7, 0.8)$ & 49 & 4 & 45 &  898 &  9975 & $0.74 \pm 0.22$ & $0.75 \pm 0.07$ \\
$[0.8, 0.9)$ & 60 & 4 & 56 &  753 & 10239 & $0.89 \pm 0.27$ & $0.91 \pm 0.07$ \\
$[0.9, 1.0)$ & 58 & 2 & 56 &  578 & 10139 & $0.58 \pm 0.25$ & $0.92 \pm 0.07$ \\
\hline
\end{tabular}
\end{table*}

\begin{figure}[htbp]
\center
\includegraphics[width=7cm,height=5.52cm]{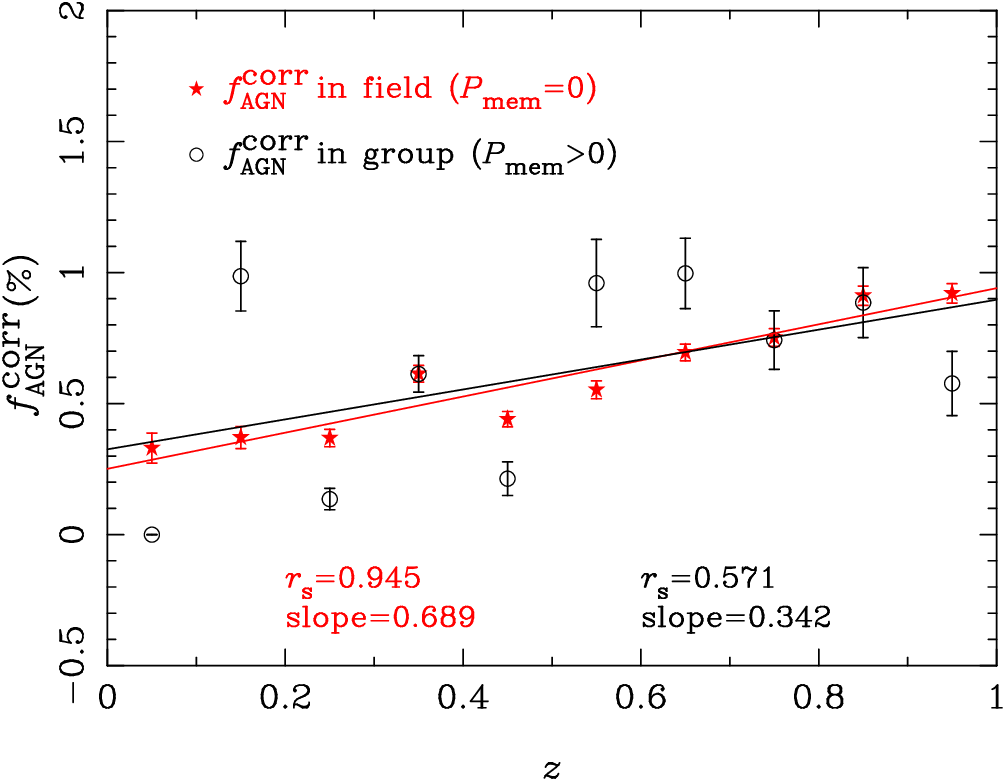}
\caption{The AGN fraction as a function of redshift in groups ($P_{\rm mem} > 0$, black circles) and fields ($P_{\rm mem} = 0$, red stars). The solid lines are the best linear fits for the AGN fraction in groups (in black) and in fields (in red).
}
\label{ffig6}
\end{figure}

For observing the probable environmental effect on AGN fraction, we investigate the AGN fraction in fields. We select the member AGNs/galaxies in groups and in fields by setting $P_{\rm mem} > 0$ and $P_{\rm mem} = 0$, respectively. The statistics of galaxies/AGNs in groups and isolated fields is presented in Table ~\ref{table2}.  The overall AGN fractions in 10 $z$-bins (i.e., $\Delta z = 0.1$) are calculated. The AGN fractions in groups and fields as a function of $z$ are shown in Fig.~\ref{ffig6}. We find that there is a strong correlation between the field AGN fraction and $z$ with a Spearman correlation coefficient of $r_s=0.945$. For the corrected AGN fraction in groups, the correlation coefficient $r_s=0.461$, which is larger than those for stricter membership selections shown in Fig.~\ref{ffig5} . The AGN fraction in isolated fields also exhibits a rising trend with $z$, and the slope is consistent with that in groups. The environment in galaxy groups seems to have no obvious influence on detection probability of the AGNs with emission features, indicating that the physical process concerning AGN fueling seems to be insensitive to the group environment. At $z>0.7$, the difference of AGN fraction between high- and low-density environments is negligible, which is consistent with the result in \citet{Mart13} that the X-ray AGN fractions in the field and clusters are similar at $1 < z < 1.5$.  However, \cite{Kauf04} found that fraction of  the local AGNs with strong [O III] emission decreases with local density, and twice as many AGNs  are found in low-density regions as in high.

\subsection{AGN fraction as a function of group richness}

\begin{figure}[htbp]
\center
\includegraphics[width=7cm,height=7.03cm]{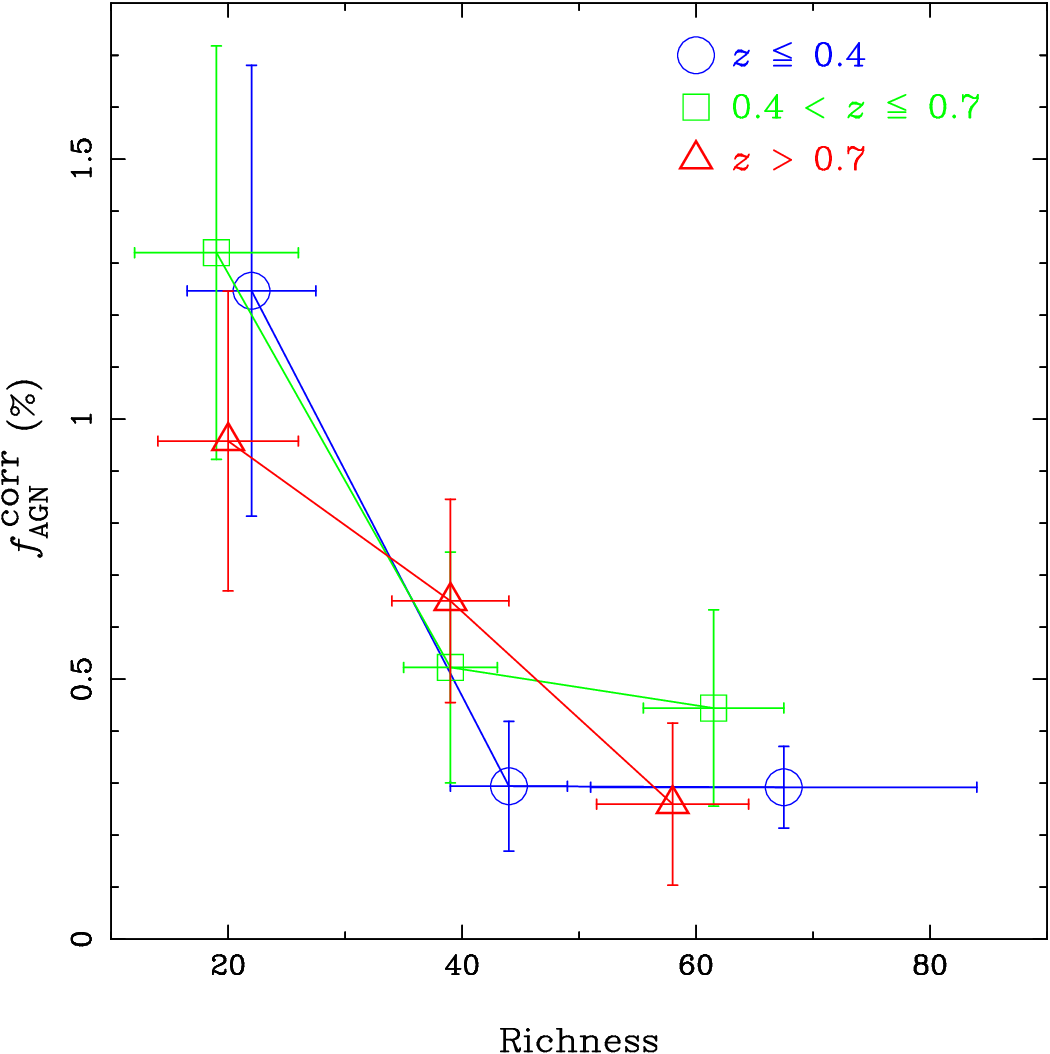}
\caption{The corrected AGN fraction as a function of group richness. Blue circles, green rectangles and red triangles denote AGN fraction in $z \leqslant 0.4$, $0.4 < z \leqslant 0.7$ and $z > 0.7$, respectively. Error bars are the median absolute deviation from median in each bin.}
\label{ffig7}
\end{figure}

To  analyze how AGN fractions vary with group richness, the samples of group galaxies/AGNs selected by $P_{\rm mem} > 0$, are split into three $z$-bins (say, $z \leqslant 0.4$, $0.4 < z \leqslant 0.7$, and $0.7 < z < 1.0$) and into three bins of  group richness (say,  $N \leqslant 30$, $30 < N \leqslant 50$, and $ N > 50$).   In Fig.~\ref{ffig7} we can see that the corrected AGN fractions show a clear decreasing trend with richness for all three redshift bins. The error bars of redshift represent the median absolute deviation from the median redshift in each $z$-bin. We find a larger fraction of emission-line AGNs in poor groups ($N \leqslant 30$), supporting the view that galaxies in groups retain larger reservoirs of cold gas to fuel AGN activity than their counterparts in clusters \citep[e.g.][]{Shen07,Geor08,Arno09,Mart09,Alle12}. \citet{Tzan14} studied {\it Chandra} X-ray point source catalogs for 9 Hickson Compact Groups (HCGs) at a median redshift $z_{\rm med} = 0.03$, and concluded that the AGN fraction in groups is higher than that in clusters.  In stark contrast to the above conclusion, some authors yielded the percentage of AGN in clusters is similar as that in fields while using different AGN selection criteria \citep[e.g.][]{Hagg10,Kle-Sar12,Mart13}.

A larger fraction of the AGNs with strong emission lines in poor groups suggests that the density environment may not only influence on the fueling mechanisms but also on the evolution of galaxies. Majority of the AGNs in poor groups are probably in the high-accretion optically dominant phase, whereas the AGN population in rich clusters is mostly in the low-accretion X-ray dominant phase. This finding is in accordance with the scenario of AGN accretion evolution that was proposed by \cite{Shen07}.

\section{Conclusions}

Based on a sample of 198 X-ray galaxy groups in a rest frame 0.1-2.4 keV luminosity range $41.3 < {\rm log}(L_{\rm X}/{\rm erg}~{\rm s}^{-1}) < 44.1$  and a sample of the spectroscopically confirmed AGNs in the XMM-COSMOS survey, we investigate the fractions of the AGNs with emission lines in galaxy groups at $z<1$ , as well as the variations with redshift and group richness. The main conclusions can be summarized as follows:

(1) For various selection criteria of member galaxies (i.e., $P_{\rm mem} > 0.1, 0.3, 0.5$), merely 27 AGNs with emission lines are found in 26 groups . Taking the AGN fraction estimated for the strictest selection criterium  $P_{\rm mem} > 0.5$ as the upper limit,  the AGN fraction  is on average less than $4.6 (\pm 1.2)\%$ for  individual groups hosting at least one AGN. The corrected overall AGN fraction for our group sample  is less than $0.98 (\pm 0.11) \%$.

(2) The normalized locations of AGNs show that all group AGNs with 0.5-2 keV luminosities less than $10^{42.5} {\rm erg}~{\rm s}^{-1})$ are found to be located at the centers of groups.

(3) The overall AGN fractions in various $z$-bins are estimated, and a week rising tendency with $z$ are found.  Overall AGN fractions in group are 0.30-0.43\% for the groups at $z<0.5$, and  0.55-0.64\% at $0.5 < z < 1.0. $. For X-ray groups at $z>0.5$. most member AGNs are X-ray bright, optically dull, which results in a lower AGN fractions at higher redshifts.

(4)The AGN fraction in isolated fields also exhibits a rising trend with $z$, and the slope is consistent with that in groups. The environment of galaxy groups seems to make no difference in detection probability of the AGNs with emission lines.

(5) For the groups in various redshift ranges, a larger AGN fractions are found in poorer groups. For  those poor groups with less than 30 member galaxies, a larger fraction of group AGNs might still be in the high-accretion phase, whereas the AGN population in rich clusters is mostly in the low-accretion X-ray-dominant phase.

\acknowledgments

We thank the referee for his/her detailed comments and suggestions. Funding for this work has been provided by the National Natural Science Foundation of China (NSFC) (Nos. 11173016, 11433005) and by the Special Research Found for the Doctoral Program of Higher Education (grant No. 20133207110006).

\nocite{*}
\bibliographystyle{spr-mp-nameyear-cnd}

\end{document}